\documentclass[superscriptaddress,showpacs,amsmath,amssymb,twocolumn,nofootinbib]{revtex4-1}

\usepackage{amsmath, amsthm, amssymb, bm, braket}
\usepackage[dvips]{graphicx}
\usepackage{color}
\usepackage{wrapfig}
\usepackage{fancybox}

\begin{document}
\title{Supplemental note for ``Two-fermion emission from spin-singlet and triplet resonances in one dimension''}

\author{Tomohiro Oishi}
\email[E-mail: ]{toishi@phy.hr} 
\affiliation{Department of Physics, Faculty of Science, University of Zagreb (Fizicki Odsjek, Prirodoslovno-Matematicki Fakultet, Bijenicka c.32), HR-10000, Zagreb, Croatia}

\renewcommand{\figurename}{FIG.}
\renewcommand{\tablename}{TABLE}

\newcommand{\bi}[1]{\ensuremath{\boldsymbol{#1}}}
\newcommand{\unit}[1]{\ensuremath{\mathrm{#1}}}
\newcommand{\oprt}[1]{\ensuremath{\hat{\mathcal{#1}}}}
\newcommand{\abs}[1]{\ensuremath{\left| #1 \right|}}

\def \beq{\begin{equation}}
\def \eeq{\end{equation}}
\def \beqa{\begin{eqnarray}}
\def \eeqa{\end{eqnarray}}
\def \Schr{Schr\"odinger }

\def \bir{\bi{r}}
\def \ubir{\bar{\bi{r}}}
\def \bip{\bi{p}}
\def \ubip{\bar{\bi{r}}}

\begin{abstract}
Time-dependent calculation has been a suitable method to 
investigate the quantum dynamical processes. 
In Ref. \cite{2018Oishi_JPG} = [T. Oishi et al., J. of Phys. G {\bf 45}, 105101 (2018)], 
we applied this method to 
the one-dimensional two-fermion tunneling in the nuclear-physics scale. 
Beside the specific results presented therein, 
some basic formalism and methods, 
which can be helpful for further discussions and developments to 
investigate time-dependent quantum systems, 
have been awaiting our description. 
This note is devoted to describe those supplemental contents. 
We do not limit the story to the nuclear physics, but keep it 
applicable to other scales and/or targets. 
\end{abstract}

\maketitle

\section{Introduction} \label{sec:intro}
Quantum-mechanical dynamics is an essential concept to 
understand various aspects of the nuclear and subatomic physics. 
Thanks to the modern development of computers, 
it is getting possible to simulate the quantum-dynamical processes 
with high accuracy. 
These processes are generally described by the 
time-development equation, 
\beq
  i\hbar \frac{\partial}{\partial t} \ket{\psi(t)} = \oprt{H} \ket{\psi(t)}, 
\eeq
where $\oprt{H}$ is the \Schr (Dirac) Hamiltonian for the 
non-relativistic (relativistic) system of interest.

In Refs. \cite{2018Oishi_JPG,2018Oishi,17Oishi,14Oishi,2012Naru}, 
we employed the time-dependent computation to investigate 
the nuclear meta-stable phenomena, e.g. two-nucleon emission. 
Besides the specific results therein, 
we have utilized the formalism and methods, 
which can be commonly useful 
for the theoretical and computational approach 
throughout the different scales or domains. 
In this note, we describe these contents 
for the time-dependent framework. 

In the following sections, 
except in Appendix, 
we assume that the time-development operator (Hamiltonian), $\oprt{H}$, is 
(i) static, 
(ii) not self-consistently dependent on the state of interest, and 
(iii) deterministic without random numbers. 
This assumption, however, cannot be suitable to several 
applications in the nuclear and atomic physics. 
For that purpose, in Appendix, we also describe some ideas 
for the case with non-static Hamiltonian. 
Also, we do not determine the specific form of $\oprt{H}$, in order 
to keep our story linkable to various interests.

\section{Continuum Expansion} \label{sec:2}
We start with the eigenstates of the Hamiltonian $\oprt{H}$. 
Those can also contain the unbound, continuum-energy states. 
Considering the degeneration, they can be formulated as 
\beqa
 && \oprt{H} \ket{E, i(E)} = E \ket{E, i(E)}, \nonumber \\
 && \Braket{E',j(E')|E,i(E)} = \delta(E'-E) \delta_{ji}. 
\eeqa
Here the eigenenergy $E$ is real, and thus, we consider the pure 
Hermite space \footnote{This assumption is in contrast to several 
theoretical methods, in which one employs the model space with 
complex eigenenergies \cite{2002Michel,2014Myo,2017Jaga}, e.g. 
Berggren space \cite{1968Berggren,1970Berggren,1993Berggren,1996Berggren}.}. 
The $i(E)$ identifies one of the degenerating states with the same energy $E$. 
However, in the following, we omit these labels for simplicity. 
We recount these labels only when it needs.

The quantum-dynamical phenomena, 
containing some particle-emitting radioactive decays 
of atomic nuclei \footnote{Those 
include, e.g. alpha decay, proton emission, and two-proton emission.}, 
can be treated as the time-development of the meta-stable state. 
Employing the eigenstates for the basis, 
an arbitrary meta-stable state, $\ket{\psi_0}$, can be expanded as 
\beq
  \ket{\psi_0} = \int \mu (E) \ket{E} dE, 
\eeq
where $\left\{ \mu(E) \right\} \in \mathbb{C}$ are the expanding coefficients. 
Thus, $\abs{\mu(E)}^2$ gives the energy spectrum. 
In Refs. \cite{2018Oishi_JPG,2018Oishi,17Oishi,14Oishi,2012Naru}, these 
coefficients were determined by ``confining potential'' 
procedure \cite{1994Serot,1998Talou,2000Talou,2004Gurv}, 
where a meta-stable state is fixed to have the similar, 
compact distribution of density to the bound state. 
The normalization is represented as 
\beq
  1 = \Braket{\psi_0|\psi_0} = \int dE \abs{\mu (E)}^2. 
\eeq
Several physical properties of $\ket{\psi_0}$ are 
characterized by the expanding coefficients, 
as we discuss in the following.

\subsection{time evolution}
Assuming $\ket{\psi_0}$ as the initial state, 
now we consider the time evolution via $\oprt{H}$. 
Because $\oprt{H}$ is static, the time-developed state 
can be trivially given as 
\beqa
 \ket{\psi (t)} 
 &=& e^{-it\oprt{H}/\hbar} \ket{\psi_0 } \nonumber \\
 &=& \int \mu (E) e^{-itE/\hbar} \ket{E} dE. \label{eq:psi_t}
\eeqa
The expectation value of $\oprt{H}$, indicated as $E_0$, 
obviously conserves during the time evolution. 
That is, 
\beqa
  E_0 &\equiv & \Braket{\psi_0|H|\psi_0} = \Braket{\psi(t)|H|\psi(t)} \nonumber \\
      &=& \int E \abs{\mu (E)}^2 dE. 
\eeqa
This conservation, of course, originates from 
that the energy spectrum should be timely invariant in the 
Hermite space \footnote{In the non-Hermite space, on the other hand, 
there is usually a leak or gain of energy.}. 
Note that, for radioactive decays of nuclei, $E_0$ corresponds to the 
mean Q value (resonance energy) 
carried out by the emitted particle(s).

The survival coefficient, $\beta(t)$, is defined as the overlap between 
the initial and the present states. 
That is, 
\beqa
  \beta(t) &\equiv & \Braket{\psi_0 | \psi(t)} \\
  &=& \int dE' \mu (E') \int dE \mu (E) \Braket{E' | e^{-itE/\hbar} | E} \nonumber \\
  &=& \int dE \abs{\mu (E)}^2 e^{-itE/\hbar}. \label{eq:Krylov}
\eeqa
Notice that $\beta(0) = 1$, consistently to the initial normalization. 
From Eq.(\ref{eq:Krylov}), one can read that 
the survival coefficient is given by the Fourier transformation of 
the invariant-energy spectrum. 
This is nothing but the ``Krylov-Fock theorem'' \cite{47Krylov,89Kuku}. 
As one of the important quantities, 
the survival probability is also given as 
\beq
  P_{\rm surv}(t) = \abs{\beta(t)}^2, 
\eeq
which physically corresponds to the decay rule of this meta-stable state. 
One important application, especially in the nuclear physics, 
is so-called ``exponential-decay rule'' and its coincidence to 
the Breit-Wigner (BW) spectrum.

\subsection{exponential-decay rule}
Exponential-decay rule is a typical aspect of nuclear radioactive processes. 
That is, 
\beq
  P(t) = e^{-t/\tau} P(0), 
\eeq
where $P(t)$ means the probability for one radioactive nucleus to survive 
with its characteristic lifetime, $\tau$. 
Indeed, one can show that the exponential-decay rule 
is equivalent to the BW distribution of the energy spectrum. 
Namely, $\abs{\mu (E)}^2$ is assumed to have the Cauchy-Lorentz form, 
whose center and full width at the half maximum (FWHM) are $E_0$ and 
$\Gamma_0$, respectively: 
\beq
  \abs{\mu (E)}^2 = \frac{1}{\pi} \frac{(\Gamma_0 /2)}{(E-E_0)^2 + (\Gamma_0 /2)^2} \label{eq:BW1}
\eeq
with $-\infty \leq E \leq \infty$ \footnote{In practical applications, however, there should be the lower limit of the energy, as we mention in the next section.}. 
Or equivalently, 
\beq
  \ket{\psi_0} = \int_{-\infty}^{\infty} dE \sqrt{\frac{\Gamma_0}{2\pi}} 
  \frac{e^{ia(E)} }{(E_0 - i\Gamma_0/2) - E} \ket{E}, \label{eq:BW2}
\eeq
where $\left\{ e^{ia(E)} \right\}$ with $a(E) \in \mathbb{R}$ are 
arbitrary phase-factors. 
If we consider the degeneration, Eq.(\ref{eq:BW1}) is modified as 
\beq
  \abs{\mu (E)}^2 = \sum_{i(E)} \abs{\mu (E,i(E))}^2 
  = \frac{1}{\pi} \frac{(\Gamma_0 /2)}{(E-E_0)^2 + (\Gamma_0 /2)^2}. 
\eeq
It is worthwhile to note the following points. 
\begin{itemize}
 \item The normalization is trivially guaranteed: 
\beqa
 && \Braket{\psi (t) | \psi (t)} \nonumber \\
 && = \int_{-\infty}^{+\infty} dE \frac{1}{\pi} \frac{(\Gamma_0 /2)}{(E-E_0)^2 + (\Gamma_0 /2)^2} = 1. 
\eeqa
 \item Next considering the Q value, however, 
how to define the first-moment value for the BW distribution is not obvious: 
one should be careful for the range of the integration. 
At this moment, we assume the isotropic infinite range with 
the central value of $E_0$. 
That is, 
\beq
  \int_I dE \equiv \lim_{R \rightarrow \infty} \int_{E_0-R}^{E_0+R} dE. 
\eeq
Thus, the first moment of the energy is identical to the Cauchy's 
principal value: 
\beqa
 & & \Braket{\psi_0 | \oprt{H} | \psi_0} = \Braket{\psi (t) | \oprt{H} | \psi (t)} \nonumber \\
 &=& \int_I dE' \mu (E') \int_I dE \mu (E) 
     \Braket{E' | \oprt{H} | E} \nonumber \\
 &=& \int_I dE' \mu (E') \int_I dE \mu (E) 
     \delta(E'-E) E \nonumber \\
 &=& \int_I dE \abs{\mu (E)}^2 E = E_0, 
\eeqa
In the following, we omit the subscript $I$. 
\end{itemize}
By substituting Eq.(\ref{eq:BW1}) into Eq.(\ref{eq:Krylov}), 
the survival coefficient is obtained from the residue at 
the pole, $E = E_0 -i\Gamma_0 /2$. 
That is, 
\beqa
 \beta(t) &=& \frac{1}{\pi} \int dE \frac{(\Gamma_0 /2)}{(E-E_0)^2 + (\Gamma_0 /2)^2} e^{-itE/\hbar} = \cdots \nonumber \\
 &=& e^{-it(E_0 -i\Gamma_0/2)/\hbar}. \label{eq:xpdr}
\eeqa
Then, the survival probability yields the well-known exponential-decay rule: 
\beq
 P_{\rm surv}(t) = \abs{\beta(t)}^2 = e^{-t/\tau}, 
\eeq
where $\tau = \hbar/\Gamma_0$ is the lifetime of this 
meta-stable state. 
This conclusion is indeed a natural product of Krylov-Fock theorem 
applied to the BW spectrum \footnote{In another example, when one starts 
with the Gaussian spectrum, $\abs{\mu(E)}^2 \propto e^{-aE^2}$, it naturally 
concludes the Gaussian-decay rule, $P_{\rm surv}(t) \propto e^{-At^2}$}.

\subsection{multi-resonance case}
It is worthwhile to mention the case, where there exist two 
modes of finite-lifetime processes. 
In this case, the survival probability should read 
\beq
 P_{\rm surv}(t) = P_{\rm surv}^{(1)}(t) \cdot P_{\rm surv}^{(2)}(t) = e^{-t/\tau_1} \cdot e^{-t/\tau_2},
\eeq
where $\tau_i=\hbar/\Gamma_i$ is the lifetime due to the $i$th mode. 
The consistent energy spectrum 
can be given as the convolution of two Cauchy-Lorentz functions. 
That is, 
\beqa
 \abs{\mu(E)}^2 &=& \int dE' L^{(1)}(E-E') L^{(2)}(E'), \nonumber \\
 L^{(i)}(x) &=& \frac{1}{\pi} \frac{(\Gamma_i/2)}{(x-E_i)^2 + (\Gamma_i/2)^2}, 
\eeqa
where $E_i$ and $\Gamma_i$ indicate the position and width, respectively, 
of the $i$th resonance. 
One can naturally extend this conclusion to the general $N$-resonance case, 
where the spectrum should be $(N-1)$-fold convolution.

\section{Practical Problems}
In practical applications, however, one often encounters 
the situation, which looks more complicated than above discussions. 
Here we mention some points worthwhile to remember. 
\begin{itemize}
 \item There should be the lower limit for 
the expansion in the continuum energy space, 
consistently to the threshold of the emission. 
Fixing it as $E=0$, we should modify Eq.(\ref{eq:BW2}) as 
\beq
 \int_{-\infty}^{\infty} dE \longrightarrow 
 \int_{0}^{\infty} dE. 
\eeq
 \item The actual energy spectra are not limited 
to the simple BW distribution, but can show 
more complicated forms. 
Even in the nuclear radioactivity, 
especially when the spectrum has the broad width, 
the BW distribution may not be suitable for practical analysis. 
\end{itemize}
Because of these two affairs, indeed, the nuclear radioactivity 
can have the deviation from the exponential-decay 
rule \cite{1961Winter,1997Wilk,2006Muga,2011Campo}. 
In such a case, the exponential-decay rule is a good approximation, 
but only for $t \simeq \tau$.


In numerical calculations, some additional affairs should be concerned. 
\begin{itemize}
 \item The model space needs a truncation by the energy cutoff, $E_{\rm cut}$. 
Of course, this cutoff should be sufficiently large to ensure 
the convergence of results. 
 \item It is often necessary to discretize the continuum-energy space. 
Thus, Eq.(\ref{eq:psi_t}) should be modified as 
\beq
 \ket{\psi(t)} = \sum_{N} \mu_N e^{-itE_N/\hbar} \ket{E_N}, 
\eeq
where $E_N \leq E_{\rm cut}$. 
In Refs. \cite{2018Oishi_JPG,2018Oishi,17Oishi,14Oishi,2012Naru}, 
we employed the radial box, $R_{\rm box}$, for discretization. 
There, we had to employ a large box to obtain the 
convergence in the time-development calculation: 
if that box was not sufficiently large, 
there should be a contamination by the reflected component at $R_{\rm box}$. 
 \item Even with the energy cutoff and continuum discretization, 
the numerical solution of the eigenstates, $\left\{ \ket{E} \right\}$, 
may need efforts, especially for three-body or more-body systems. 
Before going to the meta-stable problem, 
it is better to check the accuracy of that solution method 
by some benchmark calculations for the well-known, bound systems. 
 \item Number of basic states, $N_{\rm max}$, should 
be sufficiently large to ensue the convergence. 
In Ref. \cite{2018Oishi_JPG}, typically $N_{\rm max}=100$-$1000$. 
 \item The way to fix the initial state, 
as well as its correspondence with the experimental situation, 
should be considered carefully. 
The initial state, especially of the (multi-)particle emission, 
is usually characterized as the state, 
where the emitted particles are confined in the narrow region, and/or 
the state, which obeys the outgoing condition. 
Even with these constraints, however, there may remain an ambiguity. 
Furthermore, obtained results after the time evolution 
may significantly depend on the selected initial state. 
In Refs. \cite{2018Oishi_JPG,2018Oishi,17Oishi,14Oishi,2012Naru}, 
we employed the phenomenological procedure with the confining potential. 
This procedure has provided a good approximation for the 
nuclear radioactive processes \cite{1994Serot,1998Talou,2000Talou,2004Gurv}. 
On the other hand, we also expect that a more smart way to determine 
the initial state will be available in future. 
\end{itemize}
Notice also that some of the above issues were not seen 
in the static, bound-state problems. 
However, they turn out to be present, when one starts to 
consider the unbound, meta-stable states.

\subsection{quantum recurrence theorem}
For the original statement of this theorem, see Refs. \cite{1957Bocchi,WKPD_QR_Theorem}.

It is worthwhile to point out the link between the time-dependent calculation and
the quantum-mechanics version of the recurrence theorem.
In the following, we assume the continuum-discretized 
space: $\int dE \longrightarrow \sum_{E_N}$. 

First we consider the two-component mixture state. 
Namely, at $t=0$, 
\beq
 \ket{\psi(0)} = \mu_1 \ket{E_1} + \mu_2 \ket{E_2},
\eeq
where the normalization is also assumed:
$\Braket{\psi(0) | \psi(0)}=\abs{\mu_1}^2 +\abs{\mu_2}^2 =1$.
The time-development then gives, 
\beqa
 P_{\rm surv} (t) &\equiv & \abs{\Braket{\psi(0) | \psi(t)}}^2 \nonumber \\
    &=& \abs{\mu_1}^4 + \abs{\mu_2}^4 + 2\abs{\mu_1}^2 \abs{\mu_2}^2 \cos \Delta_{12} t,
\eeqa
where $\Delta_{12}=(E_1-E_2)/\hbar$. 
Because of the normalization, the survival probability can be 
reformulated as 
\beq
  P_{\rm surv}(t) = 1 - 2\abs{\mu_1}^2 \abs{\mu_2}^2 (1-\cos \Delta_{12} t).
\eeq
Thus, when $t=2N\pi/ \Delta_{12}$, the time-developed state 
can be equivalent to the initial state.

We next start the same story but for the three-component mixture state. 
That is, 
\beq
 \ket{\psi(0)} = \mu_1 \ket{E_1} + \mu_2 \ket{E_2} + \mu_3 \ket{E_3},
\eeq
where $\Braket{\psi(0) | \psi(0)}= \sum_{i=1-3} \abs{\mu_i}^2 =1$ (normalization).
In this case, the time development concludes that 
\beq
  P_{\rm surv}(t) = 1 - 2\sum_{(ij)} \abs{\mu_i}^2 \abs{\mu_j}^2 (1-\cos \Delta_{ij} t), \label{eq:Reccu}
\eeq
where $\Delta_{ij} = (E_i-E_j)/\hbar$, and $(ij)$ indicates 
the combination of two labels ($i \ne j$). 

Indeed, Eq.(\ref{eq:Reccu}) is valid in a 
more general case, where the initial state is 
given by the arbitrary-number superposition. 
Namely, 
\beq
  \ket{\psi(0)}=\sum_{i=1}^{N_{\rm max} >2} \mu_i \ket{E_i}. 
\eeq
In this case, of course, the summation $\sum_{(ij)}$ contains 
$N_{\rm max}(N_{\rm max}-1)/2$ terms. 
Notice also the time-reversal symmetry, $P_{\rm surv}(-t)=P_{\rm surv}(t)$, 
as long as in the Hermite model space.

From Eq.(\ref{eq:Reccu}), 
the situation looks different from the two-component case: 
there is not the simple, periodic solution for $P_{\rm surv}(t\ne 0)=1$. 
However, 
there can be still the solution of recurrence. 
To see this, we consider the gap probability, 
\beqa
 g(t) &=& 1 - P_{\rm surv}(t) \nonumber \\
      &=& 2\sum_{(ij)=(12)}^{J_{\rm max}} \abs{a_{(ij)}}^2 (1-\cos \Delta_{ij} t), 
\eeqa
where $\abs{a_{(ij)}}^2 \equiv \abs{\mu_i}^2 \abs{\mu_j}^2$ and 
$J_{\rm max} \equiv N_{\rm max}(N_{\rm max}-1)/2$. 
Because $g(t)$ is an almost-periodic function, 
there can be the time $T$ to satisfy $g(T)<\delta^2$, 
where $\delta$ is the arbitrary positive 
number \footnote{Note that, if one employs the non-Hermite space, this recurrence is not guaranteed.}. 
Indeed, one can obtain the same conclusion more easily 
by evaluating the norm of 
$\ket{\psi(t)}-\ket{\psi(0)}$ \cite{1957Bocchi,WKPD_QR_Theorem}.

\section{Summary}
In this supplemental note following Ref. \cite{2018Oishi_JPG}, 
the basic ideas and formalism of the time-dependent 
quantum-mechanical calculation have been presented. 
Several, typical aspects of the nuclear radioactive 
processes have been also discussed. 
However, we expect that 
some contents in this note can be helpful for 
other applications, and enable us to 
proceed the wholistic study on 
quantum dynamical processes.

\appendix*

\section{NON-STATIC OPERATOR}
In the main text, the time-development operator (Hamiltonian) is 
limited to be static and independent of the present state. 
However, in several applications of quantum many-body 
problems, Hamiltonian may contain, e.g. a time-dependent 
external field, and/or the many-body interaction, 
which should be determined self-consistently to the 
corresponding state, $\psi(t)$. 
In such a case, it is necessary to deal with the 
non-static Hamiltonian, 
\beq
  \oprt{H} = \oprt{H}(t,\psi(t)). 
\eeq
In this section, we summarize the basic ideas for those applications.

The general time-dependent equation is now given as 
\beq
 i \frac{\partial}{\partial t} \ket{\psi(t)} = \oprt{O}(t,\psi(t)) \ket{\psi(t)}, \label{eq:TD_1}
\eeq
where $\oprt{O}\equiv \oprt{H}/\hbar$. 
In addition, it is worthwhile to introduce the other two equations, 
which can be equivalent to Eq.(\ref{eq:TD_1}). 
Namely, 
\beq
 \ket{\psi(t+\epsilon)} = e^{-i\epsilon \oprt{O}(t,\psi(t))} \ket{\psi(t)}, \label{eq:TD_2} 
\eeq
and also, 
\beq
 \oprt{O}(t,\psi(t)) = i \frac{\partial}{\partial \epsilon} \ln \left[ \ket{\psi(t+\epsilon)} \right]. \label{eq:TD_3}
\eeq
Here $\epsilon$ means the infinitesimal time evolution. 
Equivalence between these equations is shown as follows. 
\begin{itemize}
  \item Equivalence of Eq.(\ref{eq:TD_1}) and Eq.(\ref{eq:TD_2}). 
First we expand the left-hand side (LHS) of Eq.(\ref{eq:TD_2}) 
with respect to $\epsilon$: 
\beq
 \ket{\psi(t+\epsilon)} = \ket{\psi(t)} + \epsilon \left( \frac{\partial}{\partial t} \right) \ket{\psi(t)} + \cdots \nonumber
\eeq
We next expand the exponential operator in the right-hand side (RHS) of Eq.(\ref{eq:TD_2}):
\beqa
 & & e^{-i\epsilon \oprt{O}(t,\psi(t))} \ket{\psi(t)} \nonumber \\
 &=& \left[ 1 + (-i\epsilon)\oprt{O}(t,\psi(t)) + \cdots \right] \ket{\psi(t)}. \nonumber
\eeqa
By comparing the first-order terms, we can conclude Eq.(\ref{eq:TD_1}). 
  \item Equivalence of Eq.(\ref{eq:TD_2}) and Eq.(\ref{eq:TD_3}). 
From the derivative of Eq.(\ref{eq:TD_2}), one obtains that 
\beq
 i \frac{\partial}{\partial \epsilon} \ket{\psi(t+\epsilon)}
 = \oprt{O}(t,\psi(t)) \ket{\psi(t+\epsilon)}. \nonumber 
\eeq
This is equivalent to Eq.(\ref{eq:TD_3}). 
\end{itemize}
Remember that we have not assumed the specific form of $\oprt{O}(t)$, 
in order to keep generality. 
In some applications, one often needs to employ a kind 
of iterative methods, in order to compute the time development. 
For implementation of that method, the most convenient form is probably 
Eq.(\ref{eq:TD_2}), which 
guarantees that $\psi(t+N\epsilon)$ can be obtained 
from $\oprt{O}(t+(N-1)\epsilon)$ and $\psi(t+(N-1)\epsilon)$. 
That is, 
\begin{equation}
  \begin{array}{l rcl}
  i=0:~~~        &\psi(t)            &\Longrightarrow  &\oprt{O}(t)  \\
                 &                   &\swarrow         & \\
  i=1:~~~        &\psi(t +\epsilon)  &\Longrightarrow  &\oprt{O}(t +\epsilon)  \\
                 &                   &\swarrow         & \\
  i=2:~~~        &\psi(t+2\epsilon)  &\Longrightarrow  &\oprt{O}(t+2\epsilon)  \\
                 &                   &\swarrow         & \\
                 &\vdots             &                 &
  \end{array} \nonumber
\end{equation}
Of course, $\epsilon$ should be sufficiently small compared with 
the typical timescale of the process of interest. 
On the other hand, Eq.(\ref{eq:TD_3}) can be helpful when the time-development 
operator is unknown, but only 
the time-dependent wave function is given. 
In such a case, one can infer $\oprt{O}(t)$ by 
evaluating the logarithmic derivative of $\psi(t+\epsilon)$.


\begin{thebibliography}{9}

   \bibitem{2018Oishi_JPG} T. Oishi, L. Fortunato, and A. Vitturi, J. of Phys. G {\bf 45}, 105101 (2018). 
   \bibitem{2018Oishi} T. Oishi, Phys. Rev. C {\bf 97}, 024314 (2018). 
   \bibitem{17Oishi} T. Oishi, M Kortelainen, and A. Pastore, Phys. Rev. C {\bf 96}, 044327 (2017). 
   \bibitem{14Oishi} T. Oishi, K. Hagino, and H. Sagawa, Phys. Rev. C {\bf 90}, 034303 (2014). 
   \bibitem{2012Naru} T. Maruyama, T. Oishi, K. Hagino and H. Sagawa, Phys. Rev. C {\bf 86}, 044301 (2012). 


   \bibitem{1994Serot} O. Serot, N. Carjan, and D. Strottman, Nucl. Phys. A {\bf 569}, 562 (1994). 
   \bibitem{1998Talou} P. Talou, N. Carjan, and D. Strottman, Phys. Rev. C {\bf 58}, 3280 (1998). 
   \bibitem{2000Talou} P. Talou, N. Carjan, N. Negrevergne, and D. Strottman, Phys. Rev. C {\bf 62}, 014609 (2000). 
   \bibitem{2004Gurv} S. A. Gurvitz, P. B. Semmes, W. Nazarewicz, and T. Vertse, Phys. Rev. A {\bf 69}, 042705 (2004). 


   \bibitem{2002Michel} N. Michel, W. Nazarewicz, M. P{\l}oszajczak, and K. Bennaceur, Phys. Rev. Lett. {\bf 89}, 042502 (2002). 
   \bibitem{2014Myo} T. Myo, Y. Kikuchi, H. Masui, and K. Kato, Prog. Part. Nucl. Phys. {\bf 79}, 1 (2014). 
   \bibitem{2017Jaga} Y. Jaganathen et al., Phys. Rev. C {\bf 96}, 054316 (2017). 
   \bibitem{1968Berggren} T. Berggren, Nucl. Phys. A {\bf 109}, 265 (1968). 
   \bibitem{1970Berggren} T. Berggren, Phys. Lett. B {\bf 33}, 647 (1970). 
   \bibitem{1993Berggren} T. Berggren, and P. Lind, Phys. Rev. C {\bf 47}, 768 (1993). 
   \bibitem{1996Berggren} T. Berggren, Phys. Lett. B {\bf 373}, 1 (1996). 


   \bibitem{47Krylov} N. S. Krylov, and V. A. Fock, Zh. Eksp. Teor. Fiz. 17 (1947) 93. 
   \bibitem{89Kuku} V. I. Kukulin, V. M. Krasnopolsky, and J. Horacek, {\it Theory of Resonances: Principles and Applications} (Springer-Verlag, Berlin and Heidelberg, Germany, 1989), Reidel Texts in the Mathematical Sciences, Vol. 3. 


   \bibitem{1961Winter} R. G. Winter, Phys. Rev. {\bf 123}, 1503 (1961). 
   \bibitem{1997Wilk} S. R. Wilkinson et al., Nature {\bf 387}, 575 (1997). 
   \bibitem{2006Muga} J. G. Muga et al., Phys. Rev. A {\bf 73}, 052112 (2006). 
   \bibitem{2011Campo} A. del Campo, Phys. Rev. A {\bf 84}, 012113 (2011). 


   \bibitem{1957Bocchi} P. Bocchieri, and A. Loinger, Phys. Rev. {\bf 107}, 337 (1957).
   \bibitem{WKPD_QR_Theorem} ``Poincar\'{e} recurrence theorem'' in WIKIPEDIA.
\end{thebibliography}
\end{document}